\documentclass[12pt,aps,amsmath,amssymb,nofootinbib,prd,superscriptaddress]{revtex4} 
\usepackage{bm}
\usepackage{graphicx} 
\usepackage{amsmath}
\usepackage{amsfonts,amsbsy}
\usepackage{amssymb}
\usepackage[colorlinks=true,pdfstartview=FitV,linkcolor=red,citecolor=blue,urlcolor=blue]{hyperref}

\def\be{\begin{equation}}
\def\ee{\end{equation}}
\def\bea{\begin{eqnarray}}
\def\eea{\end{eqnarray}}
\newcommand{ \mean }[1]{\left\langle #1 \right\rangle}   

\usepackage{color}
\definecolor{dgreen}{cmyk}{1.,0.,1.,0.2}        
\definecolor{orange}{cmyk}{0.,0.353,1.,0.}    

\begin{document}

\title{Chiral Magnetic Effect Task Force Report}

\author{Vladimir~Skokov~(co-chair)}
\email{vskokov@bnl.gov}
\affiliation{RIKEN/BNL, Brookhaven National Laboratory,
Upton, NY 11973}
\author{Paul~Sorensen~(co-chair)}
\email{psoren@bnl.gov}
\affiliation{Department of Physics, Brookhaven National Laboratory,
Upton, NY 11973} 
\author{Volker~Koch}
\affiliation{Nuclear Science Division, 
Lawrence Berkeley National Laboratory, 
Berkeley, CA, 94720, USA} 
\author{Soeren~Schlichting}   
\affiliation{Department of Physics, Brookhaven National Laboratory,
Upton, NY 11973}
\author{Jim~Thomas}
\affiliation{Nuclear Science Division, 
Lawrence Berkeley National Laboratory, 
Berkeley, CA, 94720, USA} 
\author{Sergei~Voloshin} 
\affiliation{Wayne State University, 666 W. Hancock, Detroit, MI 48201} 
\author{Gang~Wang}
\affiliation{Department of Physics and Astronomy, University of California, Los Angeles, California 90095, USA} 
\author{Ho-Ung~Yee}
\affiliation{Department of Physics, University of Illinois, Chicago, Illinois 60607}
\affiliation{RIKEN/BNL, Brookhaven National Laboratory,
Upton, NY 11973}

\date{\today}
\maketitle
{\bf Executive Summary: }
In 2009, the STAR collaboration at RHIC published a letter ``Azimuthal Charged-Particle Correlations and Possible Local Strong Parity Violation''.
This letter reported on a parity even observable that is sensitive to the separation of positive and negative charges along the direction of 
the magnetic field in heavy-ion collisions that could arise from the chiral magnetic effect. The discovery of the chiral magnetic effect would confirm the simultaneous existence of ultra-strong magnetic fields, chiral symmetry restoration and topological charge changing transitions in heavy-ion collisions. 
The creation of odd parity domains in hot QCD will have wide ranging implications in the literature.
Possible contributions from effects other than the chiral magnetic effect were explored
but none of the non-parity violating models considered in that letter were found to be able to describe the observed signal. Subsequently however,
model studies of background effects and discoveries related to the importance of fluctuations in the initial geometry of heavy-ion collisions have 
shown that much if not all of the original signal reported in 2009 could arise from effects unrelated to
the chiral magnetic effect.
Further measurements and
calculations have been made, many of which fit the general expectations of chiral effects but questions of interpretation still make it difficult to
uniquely identify the chiral magnetic effect as the source of the measured charge separation.

The unique identification of the chiral magnetic effect in heavy-ion collisions would represent one of the highlights of the RHIC physics program and
would provide a lasting legacy for the field. The current plan for completing the RHIC mission envisions a second phase of the beam energy scan with 
detector and accelerator upgrades carried out in 2019 and 2020 and a jet and quarkonium program with enhanced detector capabilities in 2021 and 2022.
Within this well motivated and prescribed program, if there are opportunities to make specific advances in our understanding of the chiral magnetic 
effect in heavy-ion collisions, they need to be identified and planned for now. In this report, we briefly examine the current status of the study of
the chiral magnetic effect including theory and experimental progress. We recommend future strategies for resolving uncertainties in interpretation 
including recommendations for theoretical work, recommendations for measurements based on data collected in the past five years, and recommendations
for beam use in the coming years of RHIC.
{\bf We have specifically investigated the case for colliding nuclear isobars (nuclei with the same mass but 
different charge) and find the case compelling. We recommend that a program of nuclear isobar collisions to isolate the chiral magnetic effect from background
sources be placed as a high priority item in the strategy for completing the RHIC mission. }

\newpage

In the following report,  we review theoretical progress on the topic of the chiral magnetic effect (CME) and associated effects and point out the primary theoretical uncertainties. We will then briefly discuss the experimental status including recent results on central U+U and Au+Au collisions (see Ref.~\cite{Kharzeev:2015znc} for a review). We discuss uncertainties that remain in the interpretation of the experimental data and then recommend priorities for theory, modeling, and experimental work that can be carried out with currently available data sets. Finally, we report our findings regarding the viability of using nuclear isobar collisions to make progress on resolving uncertainties related to the interpretation of the experimental measurements in terms of the CME. We find that a modest program to collide isobars at RHIC will reduce the uncertainty on the magnetic field dependence of the observed charge separation by more than a factor of five. If the CME is responsible for even a modest portion of the experimental charge separation signal ($\gtrsim 20$\%), this program will enable a statistically significant measurement of the CME. Although continued analysis of current data sets is important, a nuclear isobar program which makes it possible to independently manipulate the magnetic field is likely to be more definitive than the measurements relying on current data.

\section{Theory Progress and Summary}
\label{the_prog_and_sum}

The CME in its  most comprehensive form is a phenomenon of charge transport along a background magnetic field: $ \vec{J} = \sigma_\chi \vec{B}$. 
For a plasma in quasi-equilibrium with approximately conserved axial symmetry 
and in a static, homogeneous magnetic field, the vector and the axial-vector currents are~\cite{Fukushima:2008xe,Vilenkin:1980fu,Son:2004tq}
\be
\vec{J}={e \mu_A\over 2\pi^2} \vec{B}\,,\quad \vec{J}_A={e\mu \over 2\pi^2} \vec{B}\,.\label{theq1}
\ee
These results have been confirmed in several theoretical frameworks. 
In perturbative QCD (pQCD), the time scale of 
axial charge relaxation from either sphaleron transitions or finite quark masses are respectively of the order $\tau_R\sim (\alpha_s^5 \log(1/\alpha_s) T)^{-1}$~\cite{Bodeker:1998hm,Arnold:1998cy} or
$\sim (\alpha_s m_q^2/T)^{-1}$~\cite{Grabowska:2014efa,misha-yee}. These time scales are  long compared to the hydrodynamic scale
$\sim (\alpha_s^2 T)^{-1}$ where the above form of the CME is valid. Dynamical electromagnetism gives rise to a further mechanism of axial
charge conversion to magnetic helicity via an instability~\cite{Joyce:1997uy,Laine:2005bt,Boyarsky:2011uy,Akamatsu:2013pjd,Manuel:2015zpa,Hirono:2015rla}, but this is suppressed by $\alpha_{EM}$. Numerically these time scales in heavy-ion 
collisions are shown to be longer than and at best comparable to $10$ fm. Therefore, the CME in a QCD plasma is theoretically robust within the time scale of heavy-ion collisions.

\begin{figure*}[hbtp]
\centering\mbox{
\includegraphics[width=0.5\textwidth]{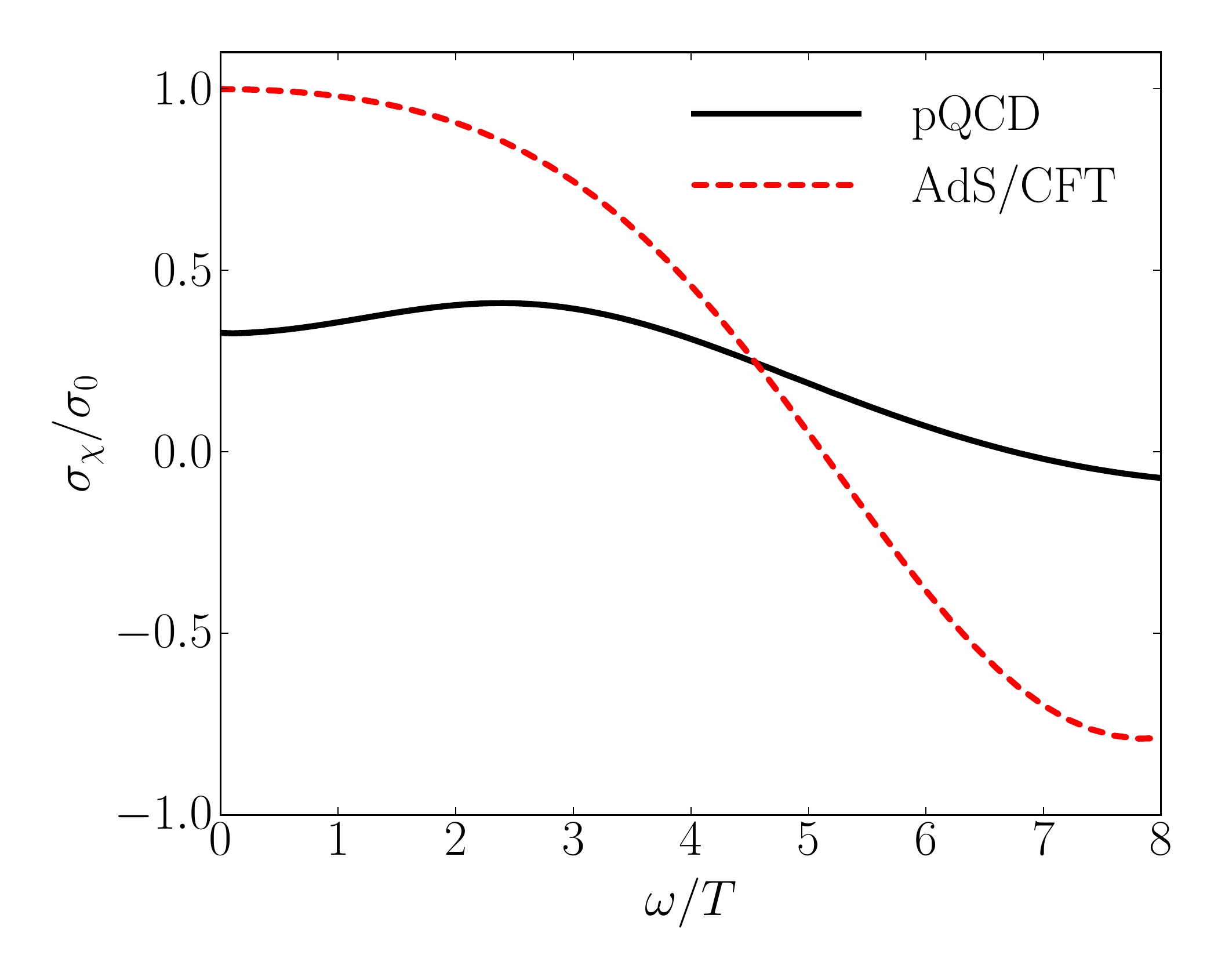}}
\caption{ 
The chiral-magnetic conductivity, $\sigma_\chi$, as a function of the frequency $\omega$ in perturbative QCD (solid line) and AdS/CFT (dashed line), normalized with respect to the static equilibrium value $\sigma_0=e\mu_A/(2\pi^2)$ in Eq.~\eqref{theq1}. 
The calculations are performed for $\mu/T=0.1$.} 
\label{f1}
\end{figure*}

Nonetheless, the time evolving magnetic field and pre-equilibrium conditions in the early times of heavy-ion collisions present challenges and have motivated recent
theoretical progress towards understanding the generalization of the original form of the CME to non-equilibrium situations. The magnitude of the CME in
non-equilibrium conditions is expected to be sensitive to the microscopic real-time dynamics of the quark-gluon plasma, which calls for reliable,
systematic theory analysis. In the weakly coupled regime, there has been recent progress on the kinetic theory of chiral fermions (``chiral kinetic theory'')
as well as in diagrammatic re-summation analysis in pQCD (see Refs.~\cite{Son:2012wh,Stephanov:2012ki,Chen:2012ca} and \cite{Jimenez-Alba:2015bia}, respectively). In strongly coupled regime, the AdS/CFT correspondence with a chiral anomaly built
in through a Chern-Simons term has provided us with useful benchmark results. Fig.~\ref{f1} shows the comparison between pQCD (solid line) and AdS/CFT 
(dashed line) of the CME conductivity $\sigma_\chi$ as a function of the frequency of the applied magnetic field (see Refs.~\cite{Kharzeev:2009pj}  and \cite{Yee:2009vw} for details). However, even given the current progress, we are by no means close
to a satisfactory understanding of the CME in non-equilibrium conditions (see Sect.~\ref{the_uncert}).

It has been proposed that an experimental signature of the CME in heavy-ion collisions may be seen in three-particle azimuthal correlations of charged
particles~\cite{Voloshin:2004vk} (see Sect.~\ref{exp_stat} and Sect.~\ref{uncert_in_interp}). These observables are sensitive to the CME charge transport along the magnetic field in the presence of axial charge.
Axial charge, which has distinctive P and CP-odd parities, can fluctuate event-by-event via topological sphaleron transitions in the early time Glasma
phase~\cite{Kharzeev:2001ev} as well as by thermal sphalerons in later stages. Understanding the magnitude and distribution of these topological fluctuations is one of the crucial difficulties
in making theoretical predictions of the proposed observables (see Sect.~\ref{the_uncert}). In the weak coupling regime, there has recently been important progress in estimating
these fluctuations by real-time lattice QCD simulations~\cite{Mace:2016svc}.

The time history and space profile of the magnetic field initially provided by charged spectators and subsequently evolved in the QGP is another critical
element for the CME predictions in heavy-ion collisions. Statistical event-by-event analysis of the initial magnetic field has been studied recently~\cite{Bzdak:2011yy,Bloczynski:2012en,Huang:2015oca,SkokovUCLA}: the proposed
experimental observable is sensitive to $\langle (eB)^2\cos(2(\psi_B-\psi_2))\rangle$ which has been shown to be correlated with centrality and 
the atomic number $Z^2$. However, the lifetime of these ``bare'' magnetic fields is less than $1$ fm. It 
has been proposed that the conductivity of the QGP in
magneto-hydrodynamics may elongate the lifetime substantially (see discussion in Refs.~\cite{McLerran:2013hla,Zakharov:2014dia,Tuchin:2014iua,Gursoy:2014aka}); 
 these studies are  based on the equilibrium value of the conductivity from
lattice QCD calculations~\cite{Ding:2010ga}. Reference~\cite{Gursoy:2014aka} indicates that the magnitude of the magnetic field  drops at least by a factor of $10^{1}$ while the lifetime may be extended considerably beyond $1$ fm. 

The CME in the later stages of the evolution can be described with ``anomalous hydrodynamics''. There have been several attempts to carry out  realistic simulations
of anomalous hydrodynamics in heavy-ion collisions to quantify the contribution of the CME to the measured charge separation (see e.g. Refs.~\cite{Hirono:2014oda,Yin:2015fca}). As the CME current is proportional to the
product of axial charge and magnetic field, a reliable calculation of the magnetic field, together with realistic simulation of anomalous hydrodynamics, is critically
needed to make reliable predictions for experiments (see Sect.~\ref{the_uncert}). 

Besides the charge separation measurements related to the CME, other charge transport phenomena originating from the same microscopic chiral anomaly have been proposed: the chiral
magnetic wave (CMW)~\cite{Kharzeev:2010gd,Newman:2005hd} and the chiral vortical effect (CVE)~\cite{Erdmenger:2008rm,Banerjee:2008th,Son:2009tf}. The CMW is a collective hydrodynamic propagation of chiral charges as a consequence of CME with a sound-like dispersion relation $\omega=\vec v_\chi\cdot \vec k$, $\vec v_\chi\propto e\vec B/T^2$, and the CVE is a charge current along the fluid vorticity $\vec\omega=\vec\nabla\times\vec v$ where $\vec v$ is a fluid velocity.
The CMW should lead to a charge-dependence of elliptic flow of hadrons that is linearly dependent on the charge asymmetry
of the QGP~\cite{Burnier:2011bf} via electric quadrupole~\cite{Gorbar:2011ya}, while CVE should induce a characteristic baryonic  charge separation~\cite{Kharzeev:2010gr,Jiang:2015cva}. These effects can potentially provide complementary experimental signals of
topological fluctuations in a QCD plasma and a macroscopic manifestation of a chiral anomaly. The measured CMW observables for pions and kaons are consistent with the trend the CMW  predicts, while no known backgrounds can explain both pions and kaons data~\cite{Bzdak:2013yla,Hatta:2015hca}. However, there exists a sizable uncertainty in the CMW prediction due to the uncertainty of the lifetime of the magnetic field.
Cross correlations between these observables may also provide further experimental tests of the effects. In the next section,  we discuss theoretical uncertainties in more detail.

\section{Theory Uncertainties}
\label{the_uncert}
As discussed in Sect.~\ref{the_prog_and_sum}, the QCD plasma created in high-energy nucleus-nucleus collisions provides a promising environment to study manifestations of 
anomaly induced transport phenomena. However, given the complexity of the space-time evolution of the system, there remain significant theoretical uncertainties
with regard to quantitative understanding of the CME in heavy-ion collisions, as well as possible backgrounds coming from more conventional mechanisms. 
Concerning the quantitative understanding of the CME in heavy-ion collisions, the dominant theoretical uncertainties originate from 
\begin{itemize}
	\item[A)] the initial distribution of axial charges,
	\item[B)] the evolution of the magnetic field, 
	\item[C)]  the dynamics of the CME during the pre-equilibrium stage,
	\item[D)] the uncertainties in the hadronic phase and the freeze-out.
\end{itemize}
 
{\bf Initial distribution of axial charges:}
There are many possible sources to generate local imbalances of axial charge densities in a QCD plasma, such as topological transitions (sphalerons), 
local fluctuations of topological density, as well as local axial currents due to the  chiral separation effect. Presently, no reliable data-validated model 
exists that comprehensively includes all these sources of axial charge density for the spatial distributions of axial charge at initial times as well
as the additional sources during the later stages of the evolution. Since the CME current is locally proportional to the axial charge density, this
constitutes a major theory uncertainty of CME in heavy-ion collisions.
 
{\bf
Evolution of the magnetic field:}
Theoretical estimates suggest that the primordial magnetic field from the spectator nucleons at the top RHIC energies can be as large as $eB\sim 10 m_{\pi}^2$~\cite{Kharzeev:2007jp,Skokov:2009qp}.
However, this initial magnetic field would exist only for a time less than 1 fm, if the effects from the created QCD medium are not taken into account. 
While the evolution of the magnetic field in the conducting QCD medium can be quite different from that in vacuum, the present theory calculations disagree
with each other on how much the conductivity of the QGP prolongs the lifetime of the magnetic field. The use of equilibrium conductivity from lattice 
QCD should be questioned for such an early pre-equilibrium stage, since the electromagnetic response of the medium depends crucially on the chemical 
formation of quarks and the space-time profile of the electromagnetic fields. These questions present major uncertainties in the current theory prediction 
of the space-time evolution of the magnetic field in heavy-ion collisions, which directly affects theoretical estimate of the experimental signatures of the CME.

{\bf
Pre-equilibrium dynamics:}
Since the magnetic field is strongest during the early stages of the heavy-ion collision, it is conceivable that a sizable CME current may be generated during 
this early pre-equilibrium stage, which would have a significant impact on the subsequent hydrodynamic evolution of the charge separation.
While real-time lattice techniques are being developed to address this problem~\cite{Mueller:2016ven}, 
theoretical description
of the pre-equilibrium stage generally remains a challenge and we do not presently have any reliable theory estimate of the CME in this early pre-equilibrium stage.
  
{\bf
Hadronic phase and freeze-out conditions:}
In addition to the major uncertainties outlined above, there also exist uncertainties in the theoretical description of how the charge separation described by 
anomalous hydrodynamics eventually translates into final state hadronic observables. One  needs to extend the present anomalous hydrodynamic modeling by including
the dissipative effects and thermal charge fluctuations, as well as hadronic interactions in the realistic freeze-out models. There are plans in the Beam Energy
Scan Theory collaboration to achieve this goal, which will be crucial to address these uncertainties. These advances in realistic hydrodynamic modeling and the 
freeze-out will also be important to quantify possible background contributions to the proposed CME observables (see Sect.~\ref{uncert_in_interp}). Since  conventional effects
such as local charge conservation and transverse momentum conservation can lead to significant background contributions to the proposed charge dependent azimuthal
correlations, the present theoretical estimates on these background effects clearly need to be improved to a modern standard of hydrodynamic modeling of heavy-ion
collisions.

While much progress has been made on the theoretical description of
chiral dynamics in heavy-ion collisions, and much progress is expected
in the near future, it is not likely that we can rely on theory alone
to arrive at robust scientific conclusions about the observation or
non-observation of CME. Having discussed the current status of theory,
we now turn to discuss the status of experimental efforts.

\section{Experimental Status}
\label{exp_stat}

The CME is expected to give rise to charge separation along the
magnetic field which in heavy-ion collisions is usually perpendicular to the reaction plane $\Psi_{\mathrm {RP}}$.  The azimuthal
distribution of particles can be expressed in the following form to
include charge separation across the reaction plane,
\be dN/d\phi \propto
1+2v_1\cos(\phi-\Psi_{\mathrm {RP}}) + 2v_2\cos[2(\phi-\Psi_{\mathrm {RP}})] +...+ 2a_\pm
\sin(\phi-\Psi_{\mathrm {RP}})+...  
\ee 
where $v_1$ and $v_2$ account for the directed and elliptic flow, and
$a_+ = -a_- \propto \mu_5 B$. Note that $\mu_5$ arising from
fluctuations takes different signs from event to event, and on event
average this dipole term vanishes, making a direct observation of this
P-odd effect impossible. Indeed the STAR measurements of
$\mean{a_\pm}$ indicate no significant charge dependence in all
centrality intervals in Au+Au collisions at 200 GeV, where the typical
difference between positive and negative charges is less than
10$^{-4}$.

What can be measured is the event-by-event correlation of $a_\pm$, a
term $\mean{a_\alpha a_\beta}$ where $\alpha$ and $\beta$ represent
electric charge. This comes at the cost of interpreting a now
P-even observable that is vulnerable to background effects. 
To suppress the background effects we make
a subtraction between the desired out-of-plane correlation and the
in-plane correlation: 
\begin{eqnarray}
&&\gamma \equiv \langle \cos(\phi_\alpha +\phi_\beta -2\Psi_{\rm RP}) \rangle =
 \langle \cos\Delta \phi_\alpha\, \cos\Delta \phi_\beta \rangle
- \langle \sin\Delta \phi_\alpha\,\sin\Delta \phi_\beta \rangle
\label{eq:obs1}
\\ && = [\langle v_{1,\alpha}v_{1,\beta} \rangle + B_{\rm IN}] -
   [\langle a_{\alpha} a_{\beta} \rangle + B_{\rm OUT}] 
   \approx - \langle a_{\alpha} a_{\beta} \rangle + [ B_{\rm IN} -
     B_{\rm OUT}], \nonumber
\end{eqnarray}
where $\Delta\phi = (\phi-\Psi_{\rm RP})$, and the averaging is done
over all particles in an event and over all events.  $B_{\rm IN,OUT}$
represent contributions from P-even background processes. This
subtraction scales down the contribution from background correlations
by approximately a factor of $v_2$. However, the CME charge separation signal
may be very small while other sources of fluctuations like
rapidity even $v_1$ fluctuations that can potentially contribute to the measurements
are known to be large.

The first measurements of the $\gamma$ correlator were made for
Au+Au and Cu+Cu collisions at 62.4 and 200 GeV with data from the
2004 and 2005 RHIC runs~\cite{Abelev:2009ac,Abelev:2009ad}.
The opposite-charge and the same-charge
correlations were found to display the ‘‘right’’
ordering, supporting the picture of the CME. The opposite-charge
correlations however were found to be close to zero or even negative
when the signature of charge separation across the reaction
plane should give rise to a positive correlation for opposite charge pairs. Measurements from the LHC exhibit very similar systematics~\cite{Christakoglou:2012mu}. These
observations are not necessarily inconsistent with CME in heavy-ion
collisions because there may be large charge-independent backgrounds
that shift the values  of both the same-charge and opposite-charge
correlations. Whether the small value of the opposite-charge
correlations is a problem or whether the difference between the
opposite-charge and same-charge correlations should be taken as the
most relevant measure is still debated.
In the early papers, background effects from conventional physics in Au+Au
collisions were simulated with the heavy-ion event generators MEVSIM,
UrQMD, and HIJING (with and without an elliptic flow after-burner
implemented). None of those generators could achieve reasonable agreement
with the data. Those generators do not however provide particularly
good descriptions of heavy-ion collisions so the fact that they don't
describe such subtle effects as charge separation provides only somewhat
limited support for a CME interpretation of the data.

  
The opposite charge correlations in Cu+Cu collisions are stronger than
those in Au+Au, possibly reflecting the suppression of the
correlations among oppositely moving particles in a larger
system~\cite{Abelev:2009ad}. STAR also presented $p_T$ and $\Delta\eta$ dependences of the
signal~\cite{Abelev:2009ad}. The signal has a $\Delta\eta$ width of about one unit of
rapidity, consistent with small P-odd domains but also narrow enough
to allow for contributions from non-CME related effects, particularly
those arising during later stages. The signal increases with the pair average
transverse momentum; it was later shown that the radial expansion can
explain such a feature~\cite{BKL}.
 The correlations from the 2007 RHIC run were measured with respect to both the
1st-harmonic plane (of spectators at large rapidity) and the
2nd-harmonic event planes at mid-rapidity. Using the first-harmonic
event plane determined by spectator neutrons ensures that the signal
is not coming from three-particle background correlations, and is due
to genuine correlations with respect to the reaction plane. Another test was
carried out by replacing one of the two charged particles 
with a neutral particle, e.g. $K^0_S$, and the results show no
signal~\cite{Zhao:2014aja}. 
Excluding pairs with low relative momenta (to suppress femtoscopic correlations) significantly
reduces the positive contributions to opposite charge correlations in
peripheral collisions, but the difference between same- and
opposite-charge correlations remains largely unchanged and consistent
with CME expectations~\cite{Abelev:2009ad}.


\begin{figure*}[hbtp]
\centering\mbox{
\includegraphics[width=0.5\textwidth]{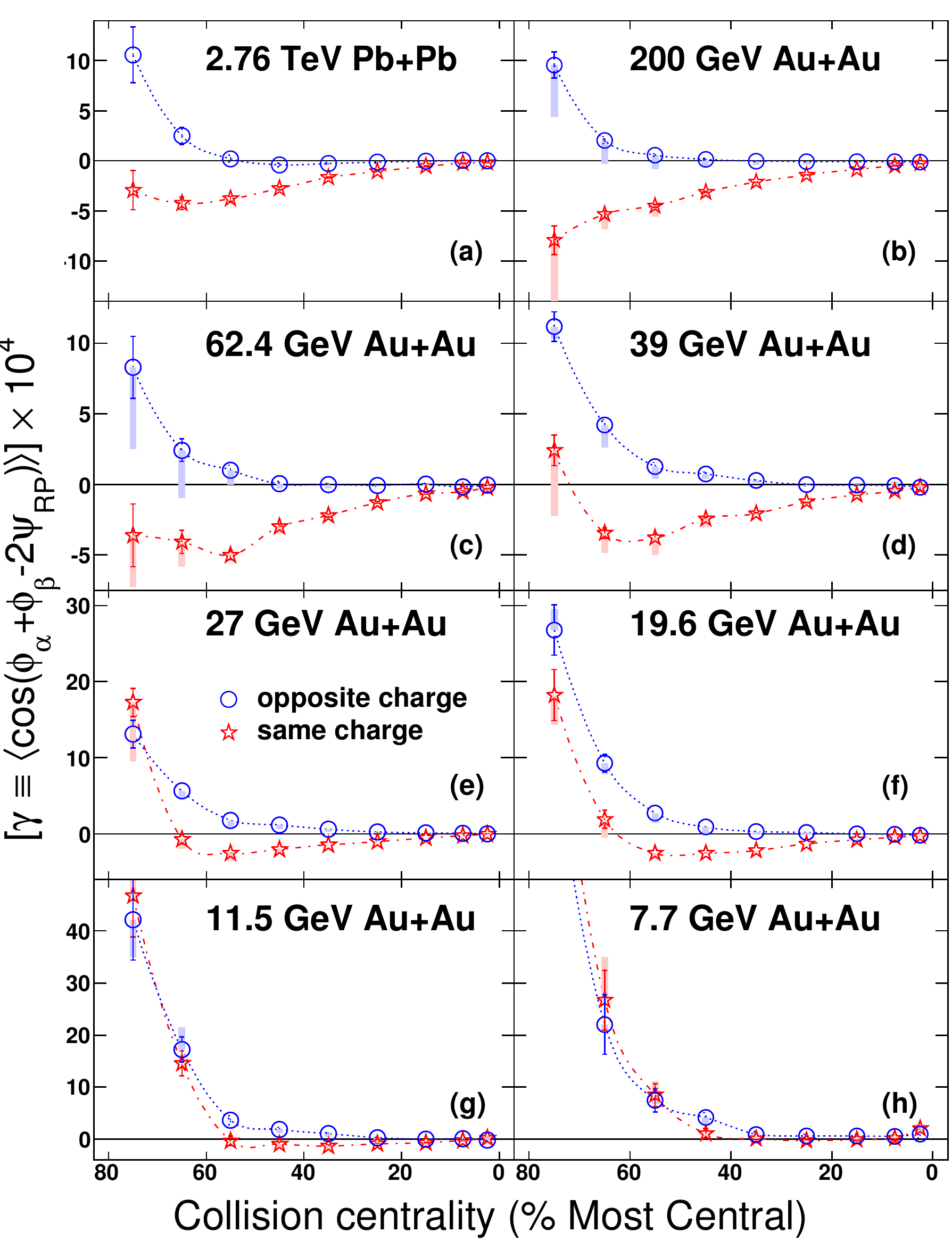}}
\caption{ 
Charge separation measurements as a function of centrality for eight collision energies spanning 7.7 GeV to 2.76 TeV.
}
\label{fbes}
\end{figure*}

The charge separation effect is thought to depend on the formation of
a quark-gluon plasma and chiral symmetry restoration. For this reason
it was speculated that the signal could be greatly suppressed or
completely absent at low collision energies where a QGP may have a
significantly shortened lifetime or may not even be formed. This could lead
to a threshold effect: with decreasing collision energy, the
signal might slowly increase with an abrupt drop thereafter. The
results from the RHIC Beam Energy Scan shown in Fig.~\ref{fbes} show
that the difference between the opposite sign and same sign correlators ($\gamma_{OS}$ and $\gamma_{SS}$ respectively) seems to
vanish at the lowest collision energy 7.7 GeV~\cite{Adamczyk:2014mzf}. At most collision energies, the
difference between $\gamma_{OS}$ and $\gamma_{SS}$ is still present
with the ‘‘right’’ ordering.
With decreased beam energy, both  $\gamma_{OS}$ and
$\gamma_{SS}$  tend to increase
starting from peripheral collisions. This feature seems to be charge
independent, and can be explained by momentum conservation and
elliptic flow. 

\begin{figure*}[!hbtp]
\centering
\begin{minipage}{0.495\textwidth}
\centering
\includegraphics[width=1.0\textwidth]{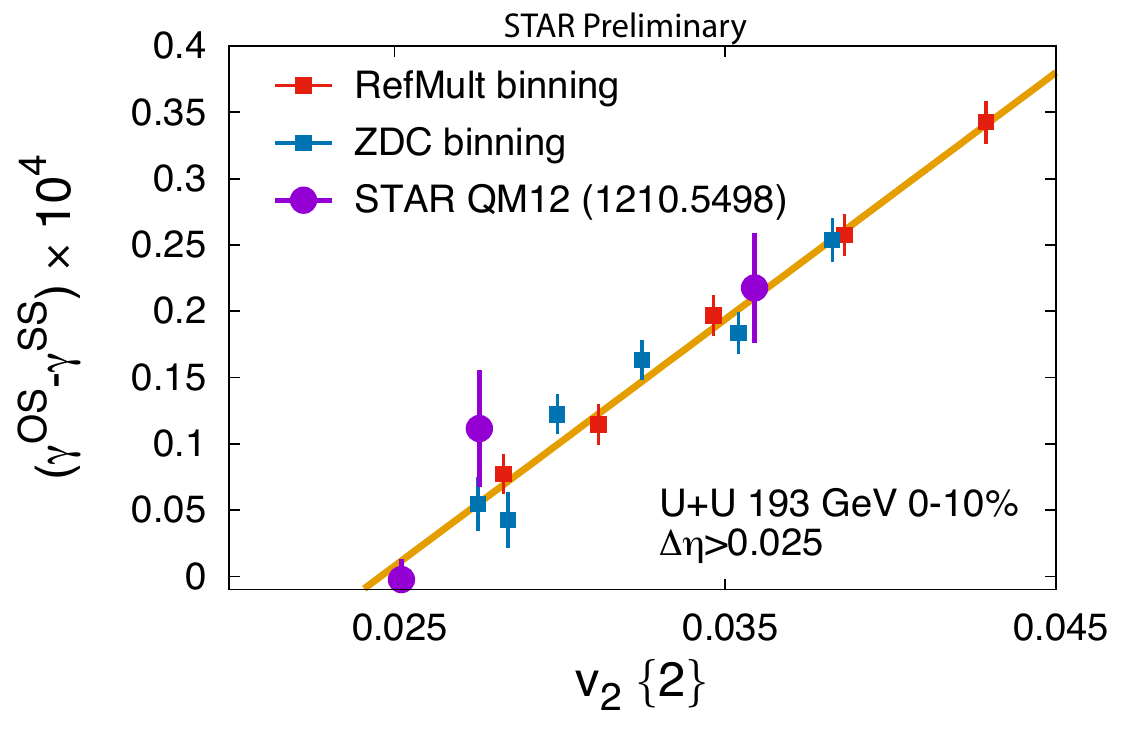}
\end{minipage}
\begin{minipage}{0.495\textwidth}
\centering
\includegraphics[width=1.0\textwidth]{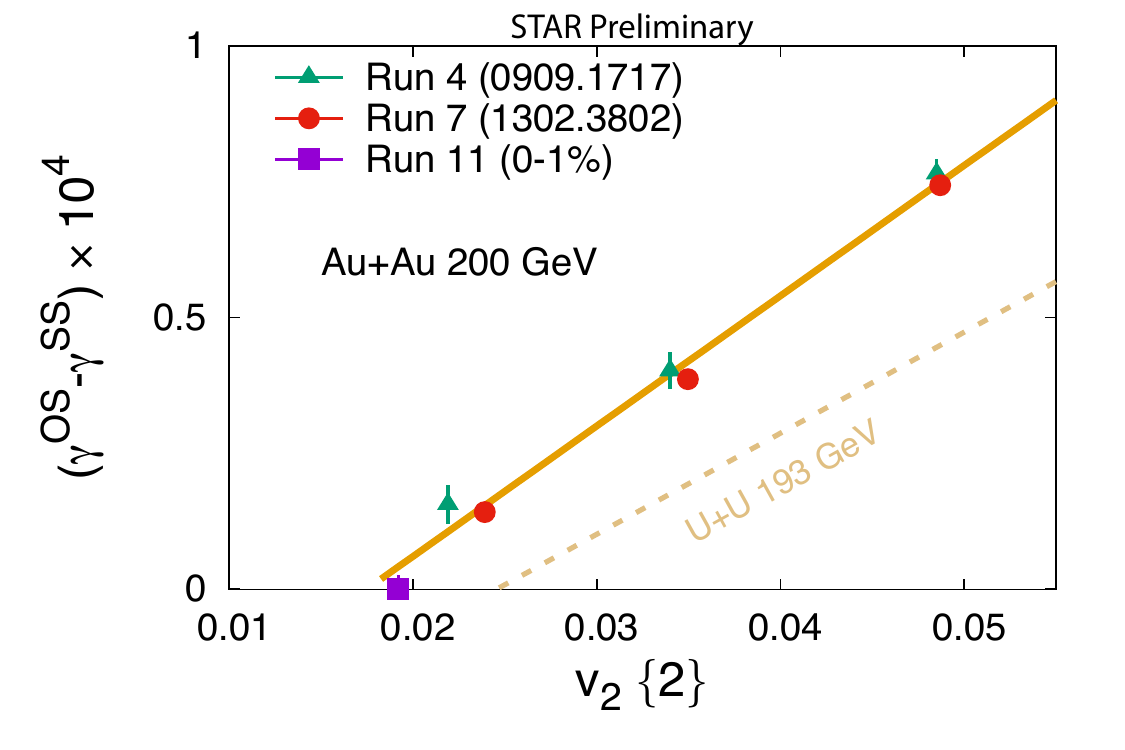}
\end{minipage}
\caption{ Charge separation as a function of $v_2$ for various centrality selections within the 0-10\% centrality range in U+U collisions (left) and Au+Au collisions (right). Preliminary results are from Ref.~\cite{TribedyUCLA}.    } \label{figuu}
\end{figure*}

Uranium nuclei have been collided at RHIC in order to study the dependence of multiplicity production, flow, and the CME on the initial overlap geometry~\cite{Kuhlman:2005ts,Voloshin:2010ut}. Early ideas focused on the idea that the prolate shape of Uranium nuclei would make it possible to select nearly fully overlapping events with large elliptic flow values but with small magnetic fields. Subsequent studies however suggest that, due to fluctuations, the square of the magnetic field is not particularly small. Measurements of very central collisions also demonstrated that the number of produced particles does not depend as strongly on the configuration of the collisions as anticipated in the two component multiplicity model leaving the experiments with a significantly reduced ability to independently manipulate the flow and the magnetic field~\cite{Adamczyk:2015obl}. Fig.~\ref{figuu} shows measurements of $(\gamma_{OS}-\gamma_{SS})$ vs $v_2$ for different centralities in
193 GeV U+U (left) and 200 GeV Au+Au collisions (right)~\cite{TribedyUCLA}. In both U+U and Au+Au collisions, the signal increases roughly with $v_2$. This initial observation may suggest that the charge separation is dominated by a $v_2$ dependence. We note however, that the charge separation goes to zero while $v_2$ is still large in central Au+Au and central U+U collisions. 
Model calculations show that the quantity $\langle (eB/m_{\pi}^2)^2\cos[2(\Psi_B-\Psi_{\rm RP})] \rangle$ as a function of eccentricity exhibits the same trend. In the models this can be traced to the fact that while $\langle B^2\rangle$ remains large due to fluctuations, $\langle \cos[2(\Psi_B-\Psi_{\mathrm RP})]\rangle$ goes to zero as $\Psi_B$ and $\Psi_{\mathrm RP}$ become decorrelated in very central collisions~\cite{Bloczynski:2012en,Chatterjee:2014sea}.
So while fluctuations in central collisions force the participant eccentricity (a positive-definite quantity) away from zero, the decorrelation of $\Psi_B$ and $\Psi_{\mathrm RP}$ drives $\langle (eB/m_{\pi}^2)^2\cos[2(\Psi_B-\Psi_{\mathrm RP})] \rangle$ to zero~\cite{TribedyUCLA}. The data therefore appear to be in better agreement with a CME interpretation than a flow background interpretation. This illustrates the importance of accounting for multiple effects including fluctuations when developing expectations for charge separation from the CME: simplistic expectations often miss important effects.

Data have also been analyzed to search for the CVE and the CMW. CMW is studied by measuring the slope $r$ of the difference between $v_2$ for positive and negative charged particles as a function of the charge asymmetry in the event $A_{\mathrm ch}=(N^+-N^-)/(N^++N^-)$: $\Delta v_2 = v_2^--v_2^+ = rA_{\mathrm ch}$. Such a slope had not been anticipated or measured prior to the prediction of its existence based on the CMW from an interplay between the CME and the chiral separation effect. Measurements of $r$ were subsequently made at RHIC across the range of energies available in the beam energy scan~\cite{Adamczyk:2015eqo} 
 and the LHC~\cite{CMW}. The measurements exhibit the trends expected from CMW. The particle-type dependence of the $\gamma$ correlator provides evidence for the existence of CVE. CVE is expected to contribute to the charge separation between baryons but not pions. In this case, measurements of charge separation for protons which are affected by both CME and CVE may be expected to be larger than for pions which are only affected by the CME. Also, measurements with neutral $\Lambda$-baryons could be sensitive to CVE but not CME. Measurements have been made of the $\gamma$ correlator to look for separation across the reaction plane for proton pairs, protons with $\Lambda$-baryons, and protons with pions~\cite{CVE}. The data exhibit an ordering p-p $\approx$ p-$\Lambda$ $>$ p-$\pi$ which is consistent with the presence of CVE with a contribution that is larger than the contribution from CME.

\section{Uncertainties in Interpretation}
\label{uncert_in_interp}

\begin{figure*}[hbtp]
\centering\mbox{
\includegraphics[width=0.55\textwidth]{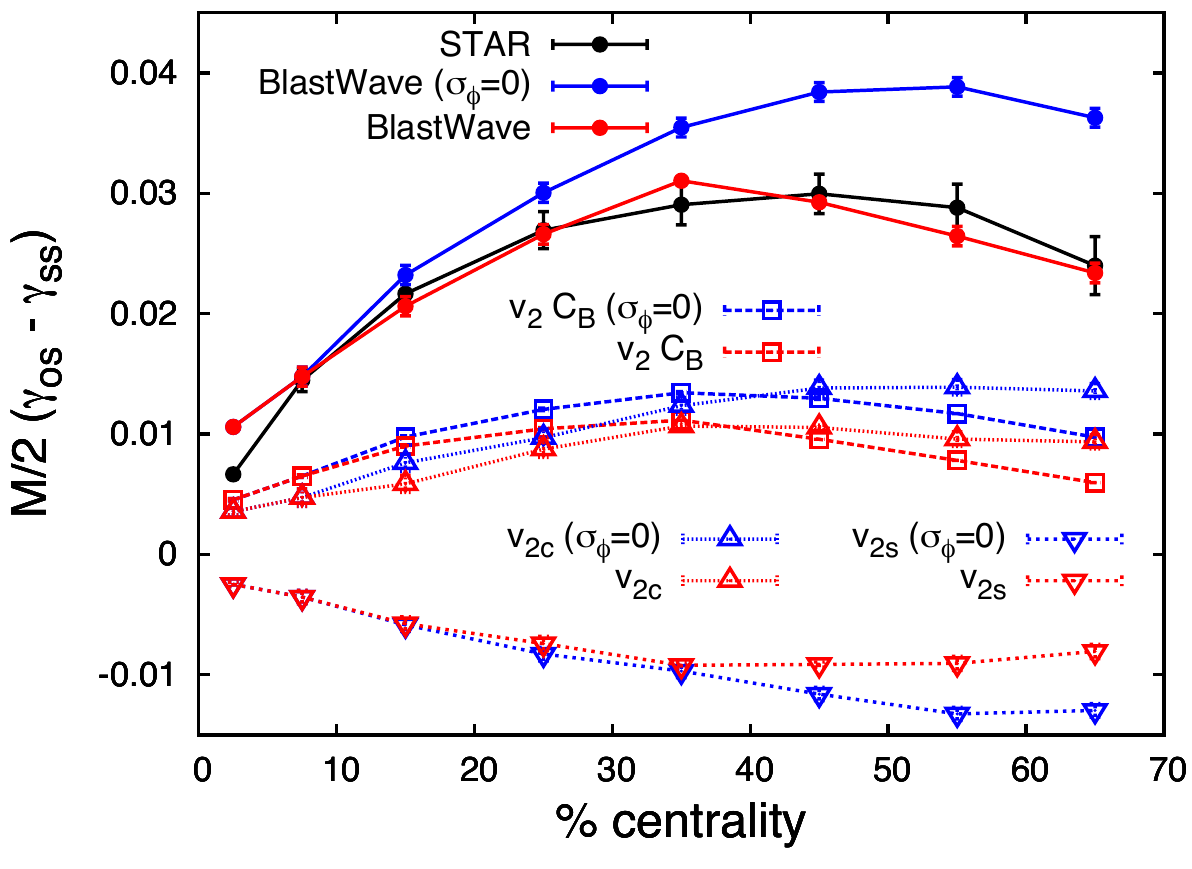}
}
\caption{Model calculations (Blast-Wave) demonstrating that the contribution of local charge conservation coupled 
with flow can mimic the observed charge separation observed by STAR~\cite{Schlichting:2010qia}. 
This indicates that background effects may account for a large fraction or all of the observed charge separation signal.
} \label{fcc}
\end{figure*}

One major source of uncertainty in the interpretation of the charge separation measurements arises from backgrounds 
induced by elliptic flow in combination with two-particle correlations. In the presence of elliptic flow, practically 
all ``conventional'' two-particle correlations contribute to the reaction-plane dependent correlation function, $\gamma$. 
Commonly discussed examples for two-particle correlation are cluster decay~\cite{Voloshin:2004vk,Wang:2009kd}, 
local charge conservation~\cite{Schlichting:2010qia},
and transverse momentum conservation~\cite{Pratt:2010zn,Bzdak:2010fd}. 
Figure~\ref{fcc}, 
for example, shows that a model that accounts for local charge conservation coupled with flow can be tuned to mimic 
the measured charge separation. Two-particle correlations also contribute to the reaction plane independent correlation, 
$\delta=\langle\cos(\phi_{\alpha}-\phi_{\beta})\rangle$, 
which will be utilized later for the background suppression in $\gamma$. 

Significant theoretical and experimental~\footnote{See e.g.~Ref.~\cite{Adamczyk:2013kcb}.} efforts have been invested into possible background
subtractions.  For example, one can inspect the $\gamma$ correlator
via the cumulant approach, where
$$
\gamma^c = \langle\langle\cos(\phi_\alpha + \phi_\beta - 2 \Psi_{\rm RP}\rangle\rangle
 =\langle\cos(\phi_\alpha + \phi_\beta - 2 \Psi_{\rm RP}\rangle -
\langle \cos[2(\phi_\beta-\Psi_{\rm RP})] \rangle\langle \cos(\phi_\alpha-\phi_\beta)\rangle 
= \gamma - v_2 \delta.
$$
This modified correlator is intended to subtract the factorizable
background contributions.  A similar, but more general correlator, has
also been introduced,
$$H = (\kappa v_{2} \delta - \gamma) /(1+\kappa v_{2}),$$ 
to suppress background contributions due to transverse momentum
conservation~\cite{Bzdak:2012ia}.  The major uncertainty in the above
expression, the coefficient $\kappa$, depends on the particle charge
combination and particle transverse momentum. It may also depend on
centrality and collision energy, reflecting slightly different
particle production mechanisms in different conditions.  For a
detailed discussion see Ref.~\cite{Bzdak:2012ia}.

The usefulness of $\gamma^c$ or $H$ can be demonstrated in a simple
and extreme example with only four particles produced in an event: a
positive particle and a negative particle go along $ \Psi_{\rm RP}$,
and another positive particle and another negative particle go to the
opposite direction. These particles show no charge separation, but
bear a strong $v_2(=1)$, and obey transverse momentum conservation and
local charge conservation.  As a result, $\gamma_{\rm SS} =
\delta_{\rm SS} = -1$ and $\gamma_{\rm OS} = \delta_{\rm OS} = 0$,
which qualitatively resemble the background-contaminated experimental
data. On the other hand,
$$
\gamma^c_{\rm OS} = \gamma^c_{\rm SS} = H_{\rm SS}^{\kappa=1}  =  H_{\rm OS}^{\kappa=1} = 0, 
$$
which reveal no charge separation.  While it is of course desirable to
eliminate background contributions to the largest extent possible, not
all of the conventional background contributions may be factorizable.
In realistic events, non-factorizable backgrounds could be
sizeable~\cite{Schlichting:2010qia}, which leaves a question mark on
whether the subtraction procedures mentioned above will lead to a
clear separation of the CME signal from the underlying backgrounds.

\section{Priorities for Theory and Modeling} 
\label{prior_for_the_and_mod}

The CME signal depends on the space-time evolution of the magnetic field as well as that of the axial charges. The latter are induced 
either by sphaleron transitions or via color flux tubes as discussed in the context of the color glass condensate. The lifetime of the magnetic 
field depends on the response of the medium to a magnetic field and may be substantially extended due to Lenz's law 
\cite{Tuchin:2014iua,McLerran:2013hla,Zakharov:2014dia,Gursoy:2014aka} depending on the electric conductivity of the medium. At the same 
time the backgrounds arise from the integral of (two-particle) correlations over the (elliptic) flow field. It has been
shown in schematic model calculations, that both signal and background are of the same order of magnitude. Therefore, a quantitative treatment
of both the signal and background is mandatory for definitive conclusions about the presence of the CME in heavy-ion collisions. This in turn
requires a quantitative understanding of the space-time evolution of the matter and its electric and magnetic properties.
  
Given the quantitative success of viscous fluid dynamics for the description of the bulk evolution of the matter created in heavy-ion collisions,
the common dynamical framework of choice is fluid dynamics, which needs to be extended in several ways:
Since  the magnetic field $\vec B$ is the driving force for anomaly-induced observables, such as the CME, it needs to be treated dynamically 
as part of the entire system evolution. In other words, one needs to solve relativistic viscous magneto-hydrodynamics (MHD). 
The presence of the triangle anomaly leads to additional sources for the various conserved currents. These can be treated by a suitable extension
of the hydrodynamic equation which is often referred to as anomalous hydrodynamics~\cite{Son:2009tf}. While first attempts have been made to 
implement this extension~\cite{Hirono:2014oda,Hongo:2013cqa,Yee:2013cya,Yin:2015fca}, a consistent treatment within a dynamical framework will be 
part of the   Beam Energy Scan Topical (BEST)  project.

In addition, it is essential to understand the beam energy dependence of the CME signal down to energies where a large part of the evolution is
governed by the dilute (non-hydrodynamic) phase. Therefore, in addition to the development of an anomalous magneto-hydrodynamic code, an equivalent
kinetic theory treatment of both the anomalous currents as well as the electro-magnetic field is necessary. Anomalous extensions for 
kinetic theory~\cite{Son:2012wh,Stephanov:2012ki,Chen:2012ca} have already been derived and need to be implemented in the treatment. 

Simultaneous to a better quantitative understanding of the CME signal, the present estimates of background effects need to be improved to a modern standard of hydrodynamic modeling of heavy-ion collisions. Since many background estimates are presently based on rather simplistic models such as blast wave parametrizations, short term progress can be achieved by incorporating these ideas into hydrodynamic models. Ultimately, it is however desirable to develop a unified framework for normal and anomalous transport based on a hydrodynamic description, including fluctuations of conserved charges, such as electric charge, baryon number and energy momentum. Similarly, also the effects of fluctuations of conserved charges on freeze-out conditions in event-by-event simulations as well as the evolution of backgrounds in the hadronic phase need to be assessed theoretically. Developments in this direction should be useful to investigate cross-correlations between different observables for CME, CMW and CVE in order to isolate common backgrounds from genuine effects.

The above program is at the center of the BEST collaboration, which has recently been selected for funding by DOE NP.

\section{Prioritized Measurements with Current Data}
\label{prior_meas_with_curr_d}

The high luminosity RHIC II upgrade envisioned in the 2007 NSAC Long
Range Plan was completed by 2010 and the STAR and PHENIX experiments
have been taking advantage of the increase in luminosity to pursue
hard probes, heavy flavor, and quarkonium studies. These large
data-sets also provide opportunities to explore the chiral magnetic
effect and related topics, differentially. In addition, RHIC has
continued to exploit its flexibility by colliding different
combinations of nuclei. These include p+Au, d+Au, He$^3$+Au, Cu+Au,
Au+Au, and U+U. These unique data-sets provide opportunities to
clarify our understanding of charge separation in heavy-ion
collisions. Not all of these opportunities have been exploited and
here we list high priority measurements that can be made with the
data-sets at hand.

While many new results have been published on charge separation, some
of the analyses either do not reveal new information beyond what was
already measured or are very difficult to interpret. We consider it
very important that new measurements satisfy the following
requirements:
\begin{enumerate} 
  \item they should be shown to be interpretable and
  \item they should either be better than previous methods, and/or
    they should provide truly new information.
\end{enumerate}
Measurements that rely on overly simplistic, semi-qualitative
arguments for interpretation should be avoided. With this in mind, we
still see promise in several measurements and point out three of them
here.

{\bf Event Shape Engineering:} Since the contribution of CME to charge
separation is expected to follow the strength of the magnetic field
while the leading background sources are expected to scale with $v_2$,
it would be useful to be able to independently vary the magnetic field
and $v_2$. 
 One of the motivations for colliding uranium nuclei was to
exploit the intrinsic prolate shape of the nucleus to collect a sample
of fully overlapping collisions with a large $v_2$ but a small
magnetic field. One of the findings from that data set is that
multiplicity production is far less dependent on the number of binary
collisions than expected so that it is harder to isolate tip-tip
collisions (small $v_2$) from body-body collisions (large $v_2$).
This significantly reduces the lever-arm available to manipulate $v_2$
in order to disentangle $v_2$ backgrounds from CME. Meanwhile, it has
also been realized that fluctuations in the magnetic field are
substantial and that the assumption that a fully overlapping collision
will have a small magnetic field is too simplistic. There are however
other ways to attempt to independently manipulate $v_2$ and the
magnetic field. Event shape engineering is one candidate that has not
been explored well enough. 
 This procedure involves selecting
events with a small q-vector~\footnote{The q-vector is defined by 
	$q = M^{-1/2} \sum_{j=1}^M \exp(2 i \phi_j)$
	where the sum is usually over all particles in a given event. For details see e.g. Ref.~\cite{Adler:2002pu}.}
as estimated from a subset of tracks in
that event, then measuring the actual $v_2$ and the charge correlators
from the other tracks.  Analyses similar to this have been pursued but
they did not account for the important step of re-estimating the $v_2$
from an independent subset of particles. In addition to event shape
engineering based on $q$-vectors, the magnetic field and $v_2$ can be
manipulated by selecting U+U collisions with asymmetric responses in
the zero-degree calorimeters. This will preferentially choose events
where the tip of one nucleus impinges on the side of the other. In
this configuration $v_2$ decreases but the magnetic field either
increases or remains unchanged~\cite{Chatterjee:2014sea}.

{\bf Higher Harmonics:} To gain more insight on
the structure of two-particle correlations with respect to the
reaction plane, the correlator $\langle \cos(m\phi_1 + n\phi_2 -
(m+n)\phi_3)\rangle$ can be measured. The usual charge separation
correlator takes $m=n=1$ so that $m+n=2$ and $\phi_3$ becomes a proxy
for the reaction plane. More detailed information can be ascertained
about the orientation of the two particle correlations with respect to
the reaction plane by examining correlators with $m=2$ and $n \ge
1$. These measurements have already been made at RHIC for inclusive
charged particles~\cite{psoren}. It should be a priority to carry out the charge
dependent measurements as well.

{{\bf Cross-correlations:} Detailed comparisons of different
  measurements or observables will also be helpful for disentangling
  the signal due to chiral phenomena from ``background''. For example,
the main suspected background contribution for both CME and CMW
measurements is local charge conservation. But one should be
able to quantitatively relate one measurement to another. Such a test
using a realistic model (the Blast-Wave model may be too simplistic) would be extremely valuable. Another example of
cross-referencing/correlation of different observables is the comparison
of the $\gamma$ correlator measurements for identified particles such as
pions, protons, and lambdas. Out of these three species, pions
``participate'' only in CME, protons in CME and CVE, and lambdas only
in CVE. As the direction of the orbital momentum and the magnetic field
are strongly aligned, the corresponding correlators should be related
one to another,
e.g. $\gamma(\Lambda,\pi) \approx 
\sqrt{\gamma(\pi,\pi)\, \gamma(\Lambda,\Lambda)}$.
}

In addition to the above, other ideas for analyses not yet performed
have been discussed. The variety and size of the data-sets now
available from RHIC-II should make it possible to carry out a number
of yet to be explored measurements. Given the complexity of a heavy-ion
collision however, it is likely that conclusive interpretations of
those measurements will require the development of a reliable model
for the chiral magnetic effect and associated backgrounds which can be
compared to the data.

\section{Recommendations for Possible RHIC Runs}
\label{rec_for_poss_runs}

In 2014 the STAR Collaboration proposed colliding nuclear isobars (pairs of nuclei with the same mass 
number A but different charge Z) as a way to vary the strength of the magnetic field while holding all 
else fixed. The proposed isobar pair was Zr+Zr and Ru+Ru. Here we investigate that proposal in more 
detail and evaluate the viability of the program. To this end we've addressed the following specific
questions:
\begin{itemize}
\item 
How well are the effects of the poorly known nuclear geometry on the measurements understood?
\item 
How well is the magnetic field understood?
\item
How well will the program be able to distinguish between background and the magnetic field dependent 
portion of the measured charge separation?
\end{itemize}

\begin{figure*}[hbtp]
\centering\mbox{
\includegraphics[width=0.99\textwidth]{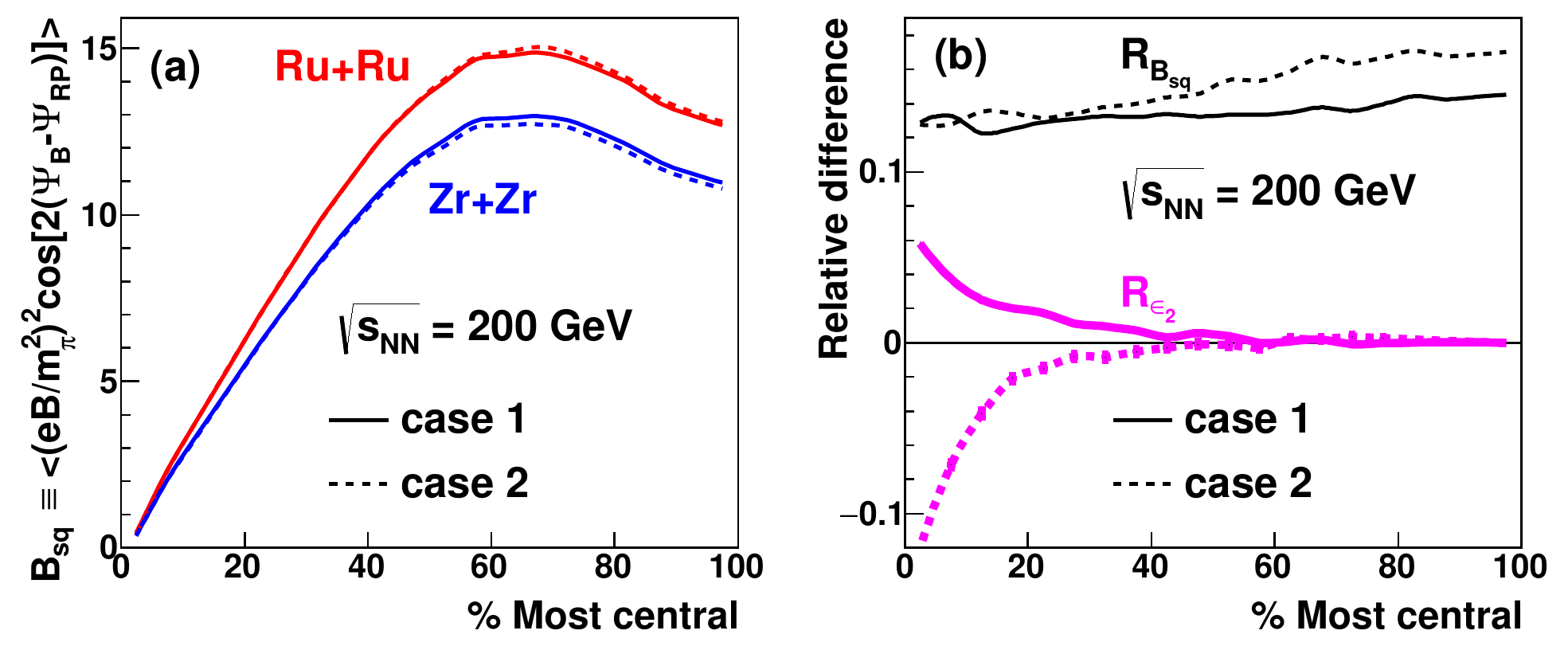}}
\caption{ Left panel: Model calculations of the square of the
  magnitude of the magnetic field and its alignment with the reaction
  plane: $\langle
  (eB/m_{\pi}^2)^2\cos[2(\Psi_B-\Psi_{\rm RP})]\rangle$. Results are shown
  for Ru+Ru and Zr+Zr collisions. Solid lines are results
  corresponding to $\beta_2$ values inferred from electron scattering
  data while dashed lines assume $\beta_2$ values from model
  calculations. Right panel: The relative difference of Ru+Ru and Zr+Zr for the two
  $\beta_2$ cases. The 10\% difference in nuclear charge translates
  into a 15\%-19\% difference in $\langle
  (eB/m_{\pi}^2)^2\cos[2(\Psi_B-\Psi_{\rm RP})]\rangle$ depending on the
  nuclear geometry. See Ref.~\cite{Deng:2016knn} for details.  } \label{fb2}
\end{figure*}

One of the central points of colliding isobar pairs is to vary the
magnetic field by a controlled amount while keeping all else
fixed. Most nuclei however are not spherical and the degree to which
they deviate from spherical varies widely. Electron scattering
measurements for example indicate that the Ruthenium nucleus has a
quadrupole deformation $\beta_2 = 0.158$ while for Zirconium the deformation
is 0.080. This then indicates that the initial geometry of a Zr+Zr
collision may be different than a Ru+Ru collision. It's important to
understand how much the shape of the nuclei will affect the ability to
draw conclusions about the magnetic field dependence of charge
separation especially since the shape of the nuclei is likely to
influence the measured $v_2$. In addition to this complication, it
appears that the shape of the nuclei is not well understood.  Model
calculations tuned to reproduce data on the shapes of other nuclei
actually predict that $\beta_2$ for Zr will be larger than for Ru
(0.217 vs 0.053) which is the opposite order compared to the electron
scattering data. As demonstrated with U+U
collisions~\cite{Adamczyk:2015obl}, a larger nuclear $\beta_2$ will
lead to a larger $v_2$ in very central collisions where the impact
parameter is small and $v_2$ is most sensitive to the shape of the
nucleus.  Measurements of $v_2$ in central Zr+Zr and Ru+Ru collisions
should therefore reliably indicate which of the two nuclei has the
larger $\beta_2$ value. Simulations using the two sets of $\beta_2$
parameters performed in Ref.~\cite{Deng:2016knn}  indicate that the ratio of $v_2$ in Ru+Ru collisions and
Zr+Zr collisions will deviate from unity by either plus 6\% or minus
12\% (see the right panel of Fig.~\ref{fb2}). These variations are
well within the sensitivity of the STAR experiment. At larger impact
parameter, the initial eccentricity of the collision overlap region is
dominated by the displacement between the nuclei and exhibits almost
no dependence on the shape of the nucleus. It's also important to
understand whether the shape of the nuclei will affect the magnetic
field. To this end, we've repeated the magnetic field calculations
while including several improvements over past calculations.

For the new calculations, instead of calculating the magnetic field at
a single point at the center of the collision system, we've integrated
over a 1 fm spot centered at the most dense region of the collision
zone. Although it's not obvious exactly what prescription should be
used for averaging over the field, this approach provides an
alternative that should account more realistically for fluctuations.
In addition, we've defined centrality classes based on estimates of
the number of produced particles instead of performing the calculation
at fixed impact parameters. As before, we perform the calculation at
time $t=0$ and use point-like protons as the charges.  These
calculations confirm that after including the effects of nuclear
geometry, the strength of the magnetic field remains proportional to
the charge Z of the colliding nuclei. Some deviation from this
behavior does arise because of fluctuations however, when calculating
the square of the magnetic field. The left panel of Fig.~\ref{fb2}
shows $\langle (eB/m_{\pi}^2)^2\cos[2(\Psi_B-\Psi_{\rm RP})]\rangle$ as a
function of centrality for Ru+Ru and Zr+Zr collisions. The solid lines
show the results when using $\beta_2$ from electron scattering data
while the dashed lines are for $\beta_2$ values estimated from model
calculations. The right panel shows the relative difference of the results for Ru+Ru
collisions over Zr+Zr collisions. We see that the
expected signal changes by between 14\% and 18\%. For the centrality
interval of interest, the expected signal from the CME varies by a few
percent depending on whether Ru or Zr is more deformed.
 
\begin{figure*}[hbtp]
\centering\mbox{
\includegraphics[width=0.99\textwidth]{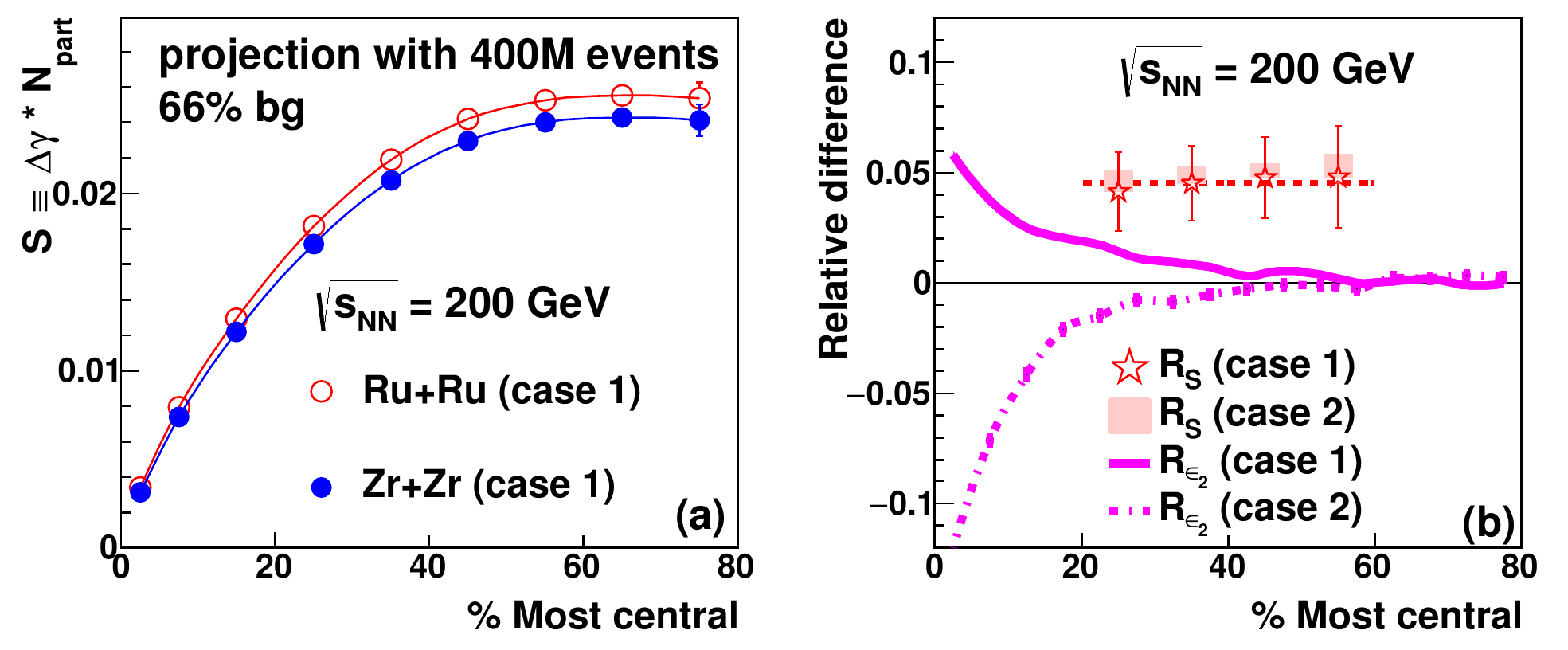}}
\caption{
Left panel: Charge separation as a function of centrality for Ru+Ru and Zr+Zr collisions assuming that 1/3 of the observed charge separation depends on the magnetic field. Error estimates assume 400 million good events observed with the STAR detector. Right panel: The relative difference between the expected charge separation along with the relative difference in the initial eccentricities for the two different possible $\beta_2$ sets.
See Ref.~\cite{Deng:2016knn} for details.
} \label{f3}
\end{figure*}

We finally turn to the question of how well the isobar program will be
able to isolate the magnetic field dependent contribution to charge
separation. We've carried out estimates based on roughly 400 million
good Ru+Ru collisions and 400 million good Zr+Zr collisions. In the
case that all of the charge separation signal is due to the CME then
$N_{\rm part} (\gamma_{OS}-\gamma_{SS})$ should be proportional to
$\langle (eB/m_\pi^2)^2 \cos(2(\Psi_B-\Psi_{\rm RP}))\rangle$. The proportionality
factor is found by plotting the experimental signal as a function of
the model calculations of $\langle (eB/m_\pi^2)^2 \cos(2(\Psi_B-\Psi_{\rm
  RP}))\rangle$ for different centralities of Cu+Cu and Au+Au collisions.
Both Cu and Au collisions seem to lie on a common curve. That curve is
then used to convert the model calculations for Ru and Zr collisions
into an expected signal. We can also account for possible
magnetic-field-independent, flow-related backgrounds by assuming some
percentage of the expected signal is independent of the magnetic
field. Fig.~\ref{f3} (left) shows the expected measurements for Ru+Ru
and Zr+Zr collisions assuming 66\% of the signal observed so far is
from magnetic field independent backgrounds.
The panel on the right
shows the relative difference between Ru+Ru and Zr+Zr along with the
relative difference in the initial eccentricity. The difference in the
initial eccentricities expected for Ru+Ru and Zr+Zr collisions is
small for mid-central collisions where the initial geometry is
dominated by the almond shaped overlap region arising at larger impact
parameters. For very central collisions where the shapes of the
colliding nuclei become more important to the initial eccentricities,
we see deviations between Ru+Ru and Zr+Zr collisions that will either
be positive or negative depending on which nucleus is more
deformed. One should be able to discern between the two cases by
measuring $v_2$ in central collisions. In either case, the effect of
the different possible $\beta_2$ sets is very small in mid-central
collisions where the magnetic field is largest. Since the geometry of
the nuclei only weakly affects the initial eccentricity and the
magnetic field, it does not seem to represent a challenge to the
success of the isobar program.

\begin{figure}[hbtp]
\centering\mbox{
\includegraphics[width=0.6666\textwidth]{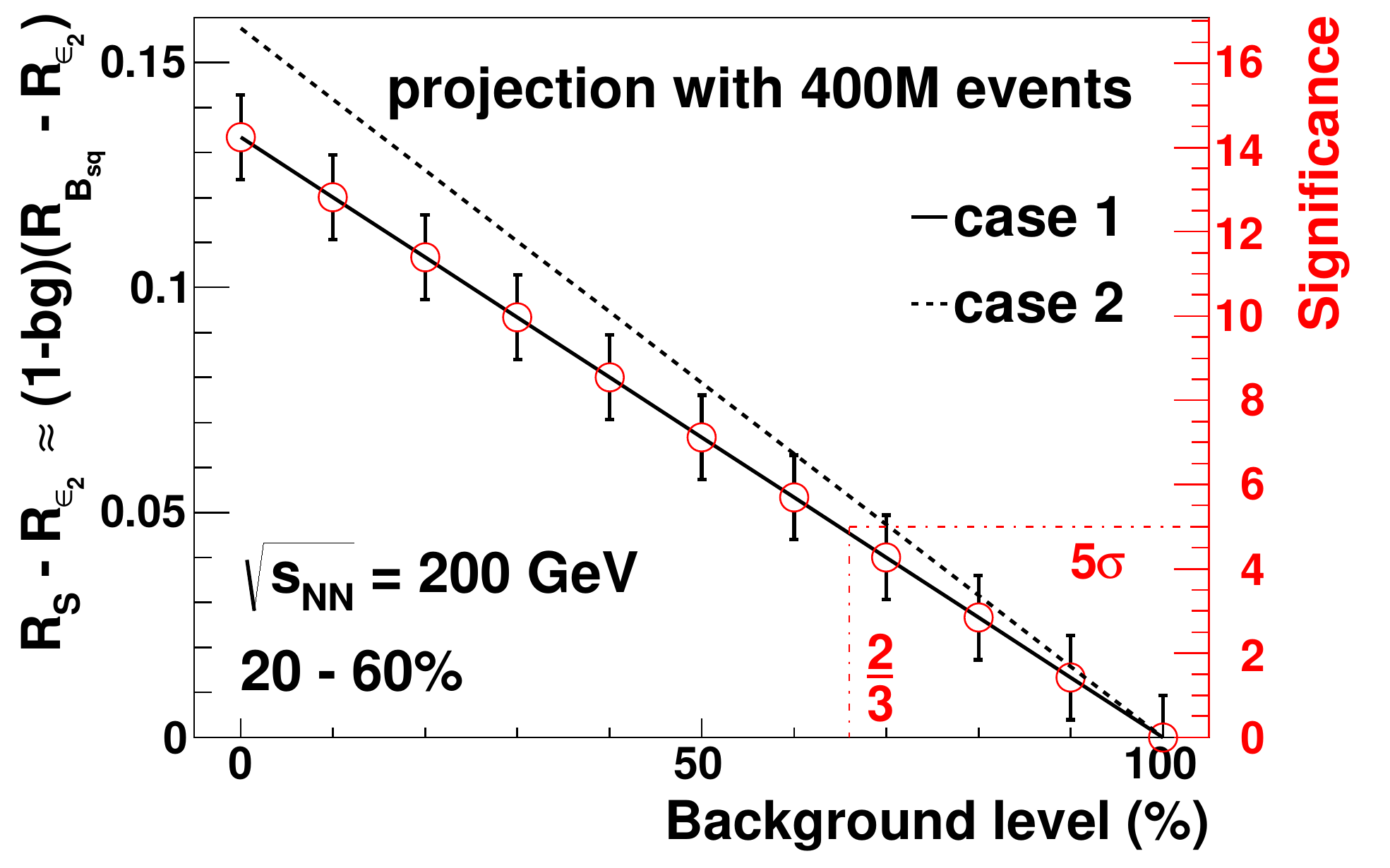}}
\caption{
The relative difference in the charge separation signal for Ru+Ru and Zr+Zr as a function of the assumed magnetic field independent background contribution with estimates of the experimental uncertainties and the equivalent significance. See Ref.~\cite{Deng:2016knn} for details.  
} \label{sig}
\end{figure}

In Fig.~\ref{sig}, we show the relative difference between Ru+Ru and
Zr+Zr collisions as a function of the assumed background level for a
$20-60\%$, centrality interval. The right axis of the figure shows the
commensurate statistical significance of the relative
difference. Assuming that all of the measured charge separation arises
from CME, we would expect an approximately 15\% difference between
Ru+Ru and Zr+Zr which for 400 million good events would be a 16-$\sigma$
effect.  We find that even if 66\% of the signal is due to background,
a 5-$\sigma$ difference due to the magnetic field dependent component of
the signal would still be observed. A three-$\sigma$ effect will still be
observed for cases where 80\% of the signal is due to background. Put
another way, this says that a program to collide Ru+Ru and Zr+Zr at
RHIC which collects 400 million good events will reduce the
uncertainty on the percentage of the background from between 0\% and
100\% to the value given by nature +/- 6.7\%. Establishing the portion
of the charge separation signal that is due to the magnetic field to
this level of precision would represent a major advance in the search
for evidence of local parity violating transitions in heavy-ion
collisions.

The change in the magnetic field in Ru+Ru and Zr+Zr collisions should
also affect the amplitude of the CMW. CMW has been studied in
experiment by measuring the slope $r$ of $\Delta v_2 = v_2^--v_2^+$ as a function of
the charge asymmetry of the event $A_{\mathrm ch}$. We find however,
that with 700 million Ru+Ru and 700 Million Zr+Zr collisions, and
assuming no magnetic field independent backgrounds, we can only expect
a one-$\sigma$ difference in $r$ between Ru+Ru and Zr+Zr collisions. An
isobar program that can be completed in one RHIC run does not appear
to provide enough statistical precision to determine to what extent
$r$ depends on the initial magnetic field.

We have also examined whether there are other isobar pairs better
suited to the purpose of these studies than Ruthenium and
Zirconium. There are three other stable sets of nuclear isobars with
charge differences of 4. These include Sn$^{124}$/Xe$^{124}$,
Te$^{130}$/Ba$^{130}$ and Xe$^{136}$/Ce$^{136}$.  For the larger
isobars, ratios of Z decrease from 44/40 for Ru/Zr to 58/54 for
Ce/Xe. We find however that the improvement expected in reaction plane
resolution from the larger multiplicities in the larger nuclei
compensates for the reduced ratio of charges so that we expect very
similar statistical significance on the measurement of the magnetic
field dependent portion of the measured charge separation. For this
reason, any of the four isobar pairs is similarly suited for the study
and the deciding factor should be the practicality of using them in
RHIC.

\section{Conclusions}
\label{conclusions}

A measurement of charge separation in heavy-ion collisions that can be
unambiguously linked to the chiral magnetic effect would be of great
interest to the wider physics community and would contribute
significantly to the scientific impact and legacy of RHIC.  Many
measurements have been carried out to study charge separation in heavy-ion
collisions that are generally in agreement with expectations from
the CME. Background models however, can also account for much of the
data. Based on our current understanding, backgrounds may account for
all of the observed charge separation. Several experimental and
theoretical steps have been identified in this report to try to
improve our understanding of the contribution of the CME to charge
separation in heavy-ion collisions. Some of these steps can be
accomplished without additional RHIC runs but we believe the most
unambiguous evidence linking the observed charge separation to the CME
is likely to come from collisions of isobaric nuclei which should make
it possible to ascertain the magnetic field dependence of the charge
separation signal. We find that such a program can reduce the
uncertainty on the magnetic field dependent contribution from 100\% to
plus or minus 6.7\%. We therefore recommend that a program to collide
nuclear isobars to isolate the chiral magnetic effect from background
sources be placed as a high priority as part of the strategy for
completing the RHIC mission.

\begin{acknowledgments}
This task force was formed by BNL Associate Laboratory Director Berndt Muller based on the advice of the 2015 Nuclear and Particle Physics Program Advisory Committee (PAC). We thank the PAC and Berndt Muller for their guidance in defining the goals of this task force. We also thank the STAR Collaboration for providing input on the proposed isobar program. We thank Jinfeng Liao for presenting a guest talk on the Chiral Vortical Effect during a committee session. We are also grateful to Dima Kharzeev and Krishna Rajagopal for useful discussions and comments. 
\end{acknowledgments}

\end{document}